\documentclass[aps,pre,showpacs,reprint,groupedaddress,amsmath,amssymb,url,floatfix]{revtex4-1}

\usepackage{epsfig}
\usepackage{color}

\definecolor{graphbach1}{rgb}{0.5,0,0}
\definecolor{graphbach2}{rgb}{0,0.5,0}
\definecolor{graphbach3}{rgb}{0,0,0.5}

\newcommand{\be}{\begin{equation}}
\newcommand{\ee}{\end{equation}} 
\newcommand{\lb}{\label}

\newcommand{\ba}{{\bf a}}

\newcommand{\br}{{\bf r}}
\newcommand{\bu}{{\bf u}}
\newcommand{\bv}{{\bf v}}
\newcommand{\bx}{{\bf x}}

\newcommand{\bW}{{\bf W}}

\newcommand{\bdot}{{\mbox{\boldmath $\cdot$}}}

\newcommand{\grad}{{\mbox{\boldmath $\nabla$}}}

\begin{document}


\title{The Diffusion Approximation in Turbulent Two-Particle Dispersion}


\author{Gregory L. Eyink${\,\!}^{1,2}$ and Damien Benveniste${\,\!}^2$}
\affiliation{${\,\!}^{1}$Department of Applied Mathematics \& Statistics\\
${\,\!}^{2}$Department of Physics \& Astronomy \\
The Johns Hopkins University, Baltimore, USA}


\date{\today}

\begin{abstract}
\noindent 
We solve an {\it inverse problem} for fluid particle pair-statistics:  we show that a time sequence  of probability 
density functions (PDF's) of separations can be exactly reproduced by solving the diffusion equation with a 
suitable time-dependent diffusivity. The diffusivity tensor is given by a time-integral of a conditional Lagrangian 
velocity structure-function, weighted by a ratio of PDF's.  Physical hypotheses for hydrodynamic turbulence 
(sweeping, short memory, mean-field) yield simpler integral formulas, including one of Kraichnan and Lundgren. 
We evaluate the latter using a spacetime database from a numerical Navier-Stokes solution for driven turbulence.  
This diffusion theory reproduces PDF's well at rms separations, but growth rate of mean-square dispersion is 
overpredicted due to neglect of memory effects.  More general applications of our approach are sketched. 
\end{abstract}

\maketitle

L. F. Richardson, in a classic paper \cite{Richardson26}, initiated the study of dispersion of particle 
pairs in turbulent flows, introducing a diffusion model 
with a scale-dependent eddy-diffusivity. There has since been much discussion about 
the accuracy of this description. In the 
case of advection by a Gaussian random velocity field which is white-noise in time, the Kraichnan 
rapid-change model \cite{Kraichnan68}, the diffusion approximation is known to be exact \cite{Falkovichetal01}. 
This fact has led to a common idea that Richardson's diffusion theory requires for its validity a 
quasi-Gaussian velocity field that is nearly delta-correlated in time \cite{Falkovichetal01,
Scatamacchiaetal12, Bitaneetal12a,Bitaneetal12b}. Nevertheless, several numerical studies have shown 
that the key predictions of Richardson's diffusion equation for the pair-separation probability density, such 
as its self-similarity in time and the precise stretched-exponential form, hold quite accurately in Navier-Stokes 
turbulence over a range of separations $0.5-2$ rms \cite{Biferaleetal05,Eyink11,Bitaneetal12b}. If Richardson's 
theory required delta-correlated Gaussian velocities, then this would be quite puzzling, because the statistics
and time-correlations of the true turbulent velocities are quite different. It is the purpose of this Letter to justify 
carefully the (limited) applicability of Richardson's diffusion theory to turbulent 2-particle dispersion. Our
approach also helps to explain deviations from Richardson's theory and to develop improved 
approximations, both topics of current interest  \cite{Scatamacchiaetal12, 
Bitaneetal12a,Bitaneetal12b}.

We exploit the following exact equation for the transition probability 
$P(\br,t|\br_0,t_0)$ of the separations of two Lagrangian particles 
in a random velocity field $\bu(\bx,t)$ that is stationary, homogeneous, and zero-mean:
\begin{eqnarray} 
&& \partial_t P(\br,t|\br_0,t_0) = \cr
&& \,\,\,\,\,\,\,\,\,\,\,\,\,\,\,\,  \,\, \partial_{r_i}\partial_{r_j} \int_{t_0}^t ds \, S_{ij}(t;\br,s|\br_0,t_0) 
P(\br,s|\br_0,t_0), \lb{Peq} \end{eqnarray} 
with conditional Lagrangian structure function 
\begin{eqnarray}
&& S_{ij}(t;\br,s|\br_0,t_0) = \int d^3x \,\, P(\bx,s|\bx_0,\br_0,t_0;\br,s) \cr
&& \times \Big\langle \delta u_i(t|\br,s;\bx)\delta u_j(\br,s;\bx)\big|\bx+\br,\bx,s; \bx_0+\br_0,\bx_0,t_0\Big\rangle.\,\,\,\,\,\,\,\,
\lb{S} \end{eqnarray}                       
Here $\bu(t|\bx,s)$ is the velocity at time $t$ of the fluid particle at $\bx$ at time $s,$ and 
$\delta\bu(t|\br,s;\bx)=\bu(t|\bx+\br,s)-\bu(t|\bx,s)$ is the Lagrangian velocity increment in label space.
The average in (\ref{S}) is conditioned upon the two particles starting at $\bx_0,\bx_0+\br_0$
at time $t_0$ and ending at $\bx,\bx+\br$ at time $s$. Likewise, $P(\bx,s|\bx_0,\br_0,t_0;\br,s) $ is the transition 
probability for a single particle starting at $\bx_0$ at time $t_0$ to arrive at  $\bx$ at time $s,$ conditioned
on a second particle starting at $\bx_0+\br_0$ at time $t_0$ and arriving at  $\bx+\br$ at time $s.$ 
The derivation of (\ref{Peq}) will be given elsewhere, but we note here that it is just a slightly more 
complicated version of Taylor's argument \cite{Taylor21} to derive an exact formula for the 
1-particle eddy-diffusivity. See also \cite{Kraichnan66, Lundgren81}. Note (\ref{Peq}) can be 
made to appear as a ``diffusion equation'' 
\be
\partial_t P(\br,t|\br_0,t_0)=\partial_{r^i}\partial_{r^j} \left[K_{ij}^*(\br,t;\br_0,t_0) 
P(\br,t|\br_0,t_0)\right], 
\lb{Preleq2} \ee
by introducing $K_{ij}^*(t,t_0)=  \int_{t_0}^t ds\,\, S_{ij}(t,s)\frac{P(s)}{P(t)}$ as an 
effective $P$-dependent diffusivity, but (\ref{Preleq2})
remains a rather complicated integro-partial-differential equation. 

By space-homogeneity, the integrand in (\ref{S}) depends only on the relative coordinate $\bx-\bx_0$
not $\bx$ and $\bx_0$ separately. It has been pointed out \cite{ThomsonDevenish05} 
that for  synthetic models of Eulerian turbulence the conditional two-time correlation must have significant dependence on 
$\bx-\bx_0$ or, alternatively, on the mean velocity $\bar{\bu}=(\bx-\bx_0)/(s-t_0)$ over time-interval $[t_0,s].$
In such models the particles are swept rapidly with velocity $\bar{\bu},$ but the turbulent 
eddies are not themselves swept. This leads to a fast decorrelation of the relative velocity of the pair, 
at a rate proportional to $\bar{\bu},$ as it is swept through the non-moving small-scale eddies. 
In Navier-Stokes turbulence, on the other hand, the eddies are advected together 
with the particles and the relative velocity of the pair should decorrelate on the slower
scale of the turnover-time of the smallest eddy that contains them. Since this set 
of eddies remains the same for any velocity $\bar{\bu}$ of the pair, there should be no dependence of 
the conditional average in (\ref{S}) upon $\bx-\bx_0$. In that case, we can integrate over $\bx$ to obtain 
\be S_{ij}(t;\br,s|\br_0,t_0) = \langle \delta u_i(t|\br,s)\delta u_j(\br,s)|\br,s;\br_0,t_0\rangle. 
\lb{S2} \ee
The considerable reduction in complexity of 
(\ref{S2}) compared with (\ref{S}) depends upon the nontrivial {\it sweeping properties} 
of Navier-Stokes turbulence. 

A straightforward simplification occurs for  $t-t_0\ll \tau_{r_0},$ where $\tau_r=r/\delta u(r)$ 
is the eddy-turnover time at separation $r.$ Taylor expansion 
about $t=t_0$ in (\ref{Peq}) gives
\be \partial_t P(t)=(t-t_0) \partial_{r^i}\partial_{r^j} \left[S_{ij}(\br_0) \delta^3(\br-\br_0)\right] +O((t-t_0)^2) 
\lb{Preleqexp} \ee
with $S_{ij}(\br)$ the usual velocity structure-function tensor. This is an exact result to leading order 
for $t-t_0\ll \tau_{r_0},$ corresponding to the Batchelor regime of ballistic separation of particles \cite{Batchelor50}. 

Deeper simplifications occur in the long-time limit $t-t_0\gg \tau_{r_0}.$ 
Note that the correlation function (\ref{S2}) is expected to decay in a time $t-s$ of order $\tau_r=r/\delta u(r)$,
while the solution $P(\br,s)$ is expected to change at a slower rate. For example, self-similar solutions 
of the type derived by Richardson have the form $P(\br,t)=L^{-3}(t)F(\br/L(t))$ with 
$L(t)\sim (t-t_0)^p$ for some power $p,$ and no dependence on $\br_0,t_0$ at sufficiently long times. 
Since $L(t)/\dot{L}(t)=t-t_0\simeq t,$ the time-scale for an order one change in $P(\br,t)$ is the current 
time $t$ for any $\br$. Thus, one should be able to substitute $P(\br,s)/P(\br,t)\simeq 1$ in $K_{ij}^*(t,t_0)$ for those $\br$ 
with $\tau_r\lesssim t.$ This {\it short-memory approximation} yields a $P$-independent ``diffusion tensor'' 
\be
K_{ij}(\br,t; \br_0,t_0)=  \int_{t_0}^t ds\,\, S_{ij}(t;\br,s|\br_0,t_0), \,\,\,\,\,\,  \tau_r\lesssim t
\lb{Kdef} \ee
Note that this approximation is reasonable for 
separations $r$ with a sufficiently rapid decay of velocity correlations, but it does {\it not} 
assume delta-correlation in time. Similarly, one can argue that the $s$-dependence 
through the conditioning event in (\ref{S2}) is slow, and approximate
\be
S_{ij}(t,s)\simeq \langle \delta u_i(t|\br,s)\delta u_j(\br,s)|\br,t;\br_0,t_0\rangle. 
\lb{Sdef2} \ee
inside the time-integral (\ref{Kdef}) defining $K(t,t_0).$ Kraichnan and Lundgren \cite{Kraichnan66,Lundgren81} 
went further in their earlier derivations and assumed (implicitly) that the $s$-dependence in Lagrangian 
particle labels also is slow. Taking $\bu(\bx,s)=\bu(s|\bx,s)\simeq \bu(s|\bx,t)$, $\bu(t|\bx,s)\simeq 
\bu(t|\bx,t)=\bu(\bx,t)$ yields
\be S_{ij}^{{\,\!}_{{\,\!}^{{KL}}}}(t,s)\simeq \langle \delta u_i(\br,t)\delta u_j(s|\br,t)|\br,t;\br_0,t_0\rangle. 
\lb{SKL} \ee

The ``diffusion equation'' (\ref{Preleq2}) with diffusivity  $K_{ij}(t,t_0)$ given by 
(\ref{Kdef}) is valid in the short-time limit $t-t_0\ll \tau_{r_0}$ also, where it reproduces the exact 
result (\ref{Preleqexp}). It is not Markovian in that limit, however, because the ``diffusion constant'' (\ref{Kdef}) is dependent 
on $\br_0,$ the separation at the initial time $t_0.$  There is strong dependence upon $\br_0$ because 
$(\br-\br_0)/(t-t_0)$ determines the relative velocity $\bv$ of the pair, which is nearly unchanging for short times.  
However, for $t-t_0\gtrsim \tau_{r_0}$ one can expect that the diffusivity becomes independent of $\br_0,t_0.$ 
More specifically, one can argue that the conditioning on the event $\{\br,t;\br_0,t_0\}$ in the average (\ref{Sdef2}) 
becomes irrelevant if $\br$ is a ``typical'' separation at time $t$, with $|\br|\simeq {\langle r^2(t)\rangle}^{1/2}.$ That is, 
for such typical separations the restricted ensemble is representative of the entire ensemble and the average may 
be evaluated without the condition:
\be S_{ij}(t,s)\simeq \langle \delta u_i(t|\br,s)\delta u_j(\br,s)\rangle, \,\,\,\, |\br|\simeq {\langle r^2(t)\rangle}^{1/2} 
\lb{Sdef3} \ee
We shall refer to this as the {\it mean-field approximation}, because it ignores fluctuations effects in separation 
$\br$ (within the stated limits). Notice when Richardson's law $\langle r^2(t)\rangle\sim g\varepsilon t^3$ 
holds, then condition $r\lesssim \langle r^2(t)\rangle^{1/2}$ coincides with the condition $\tau_r\lesssim t$
in (\ref{Kdef}) for $\tau_r=\varepsilon^{-1/3}r^{2/3}.$ However, in addition to avoiding unusually large 
separations $r\gg \langle r^2(t)\rangle^{1/2}$ one must also in (\ref{Sdef3}) avoid unusually small separations 
$r\ll \langle r^2(t)\rangle^{1/2}.$ Both of these conditions can be expected to alter the statistics of velocity increments
substantially.  Notice that a similar mean-field approximation may be made in (\ref{SKL}), yielding the 
Kraichnan-Lundgren (KL) formula for the eddy-diffusivity \cite{Kraichnan66,Lundgren81}. In either case, 
a Markovian diffusion equation is obtained for evolution of the probability distribution of pair-separations in the 
range $r \simeq \langle r^2(t)\rangle^{1/2}$. 

\begin{figure}[!t] 
\begin{center}  
\includegraphics[width=\linewidth]{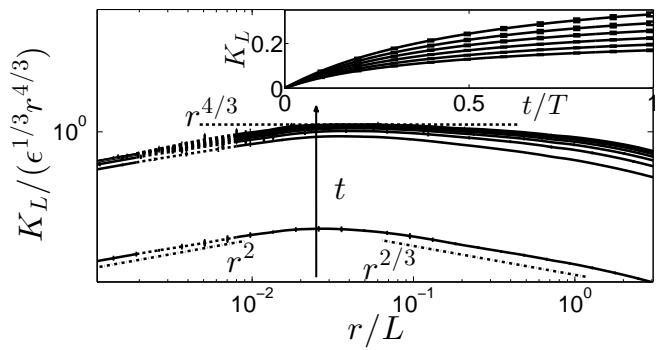}
\caption{$K_L(r,t)$ vs. $r$ for 8 different times $t=\{0.02, 6.74  , 13.46,  20.17,  26.89,  33.61,  40.33 , 44.79\}t_\nu$. The arrow indicates increasing time. 
Inset:  $K_L(r,t)$ vs. $t$ for 6 values of $r$ around $r\sim 5\times10^{-2}L$.} 
\label{diffufigure} 
\end{center} 
\end{figure}

\begin{figure}[!t] 
\begin{center}  
\includegraphics[width=\linewidth]{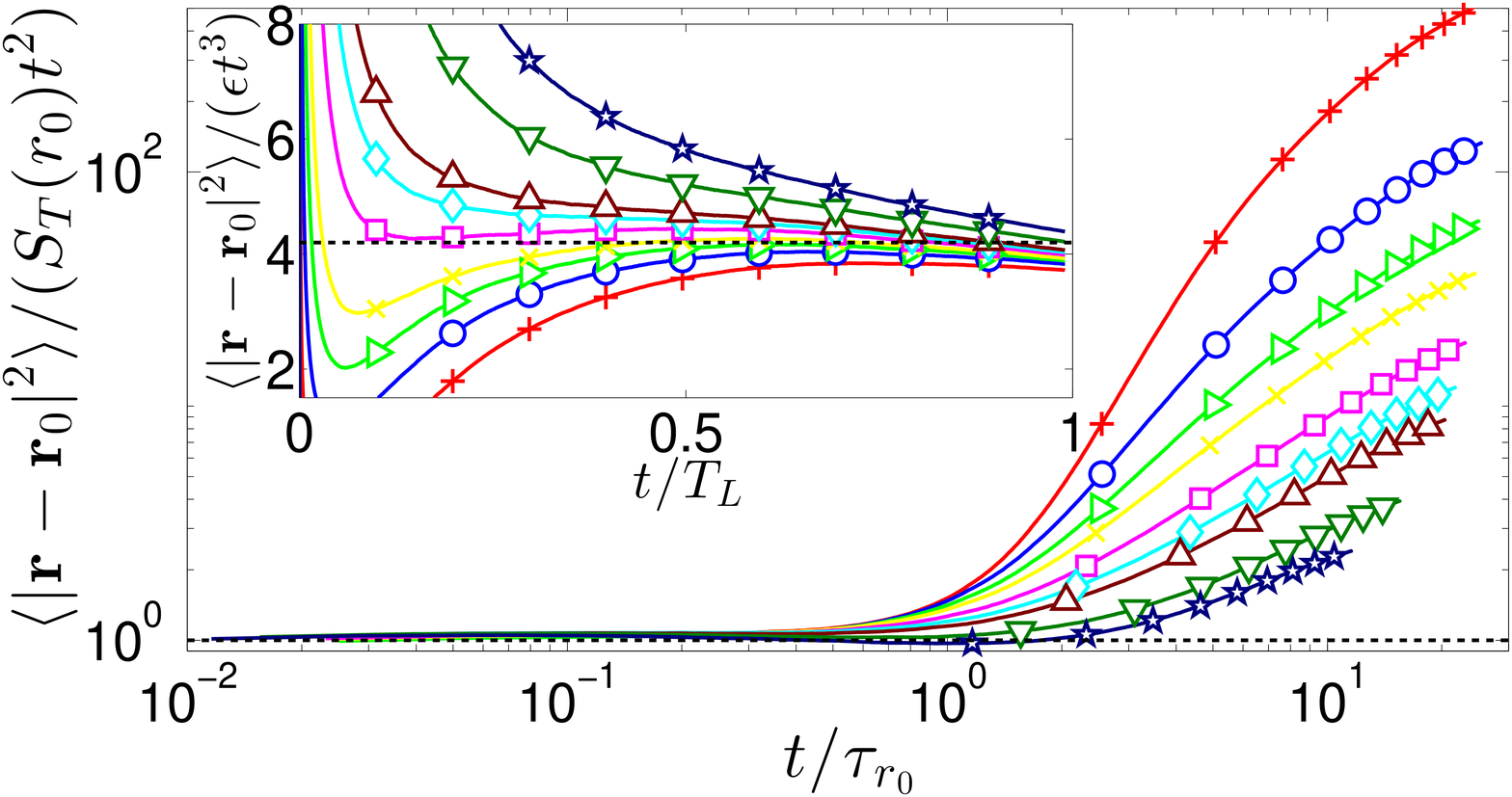}
\caption{(Colors online) Particle dispersion  $\langle|\br(t)-\br_0|^2\rangle$ for KL diffusion model compensated by the Batchelor $t^2$-law, with initial separations $r_0=l_\nu$ ($\textcolor{red}{+}$), $r_0=2l_\nu$ ($\textcolor{blue}{\circ}$), $r_0=3l_\nu$ ($\textcolor{green}{\triangleright}$), $r_0=4l_\nu$ ($\textcolor{yellow}{\times}$), $r_0=6l_\nu$ ($\textcolor{magenta}{\square}$), $r_0=8l_\nu$ ($\textcolor{cyan}{\diamond}$), $r_0=10l_\nu$ ($\textcolor{graphbach1}{\triangle}$), $r_0=20l_\nu$ ($\textcolor{graphbach2}{\triangledown}$), $r_0=35l_\nu$ ($\textcolor{graphbach3}{\star}$). Inset: curves compensated by $t^3$.\\}
\vspace{-15pt} 
\label{bachelor} 
\end{center} 
\end{figure}

In order to test validity of the physical approximations and to obtain concrete, quantitative results,  we may evaluate the 
theoretical formulas for eddy-diffusivities derived above, both exact and approximate, using turbulence data from numerical simulations. 
We here evaluate the Kraichnan-Lundgren \cite{Kraichnan66,Lundgren81} formula for the case of turbulence which is statistically stationary:
\begin{eqnarray}
K_{ij}^{{\,\!}_{{\,\!}^{{KL}}}}(\br,t)&=&\int^0_{-t} ds\,\,\langle \delta u_i(\br,0)\delta u_j(s|\br,0)\rangle. \label{diffu}
\end{eqnarray}
This formula involves pairs of particle trajectories integrated backward in time $s$ from positions displaced by $\br$ at the current time $0.$ 
It thus requires spacetime data for turbulent velocity fields.  We exploit here the JHU Turbulence Database Cluster \cite{Database1,JHUTD}, 
which provides online data over an entire large-eddy turnover time for isotropic and homogenous turbulence at Taylor-scale Reynolds number 
$Re_\lambda=433.$ The integration of particle trajectories is performed inside the database using the {\it getPosition} 
functionality \cite{Yuetal12}. Because of isotropy and incompressibility, the  diffusivity is fully defined 
by its longitudinal part $K_L(r,t)$ as a function of $r=|\br|$. The formula (\ref{diffu}) is used directly for $4\ell_\nu<r<L,$ 
with $L$ the integral scale and $l_\nu$ the Kolmogorov scale, by averaging over $N\sim6\times10^9$ particle pairs  distributed 
throughout the flow domain and integrating in $s$ by the composite trapezoidal rule. For smaller $r,$ spatial intermittency makes the ensemble average
converge slowly in $N$ and we instead expand the velocity increments in (\ref{diffu}) to leading order in $\br$ to obtain 
$K_L(r,t) =\lambda_L(t) r^2$ with 
\begin{eqnarray}
\lambda_T(t) &=&\frac{1}{3}\int_{-t}^0ds\langle\frac{\partial u_i}{\partial x_j}(\br,0)\frac{\partial u_i}{\partial x_j}(s|\br,0)\rangle \label{lamb}
\end{eqnarray}
and $\lambda_L(t)=\lambda_T(t)/5$ by incompressibility. Cf. \cite{Brunketal97}. Eq.(\ref{lamb}) can be evaluated accurately by averaging
over only $N=2\times10^4$ single-particle trajectories and the diffusivity for $r>4l_\nu$ from (\ref{diffu}) is then spline interpolated to the $r\rightarrow0$ result from (\ref{lamb}). 

Fig.~\ref{diffufigure} plots our results for $K_L(r,t)$ versus $r,$ compensated by $\epsilon^{1/3}r^{4/3},$ for 8 different times $t\in [t_\nu,T_L],$
with $t_\nu$ the Kolmogorov time and $T_L$ the large-eddy turnover time. The dashed portion 
of the curves show the interpolated range. Error bars are calculated by the maximum difference between two subensembles of $N/2$ samples. Both 
dissipation range scaling $K_L(r,t) \propto r^2$ and short-time Batchelor ballistic range scaling $K_L(r,t)\simeq S_L(r)t,$ which follow analytically from (\ref{diffu}),
are observed. For large times the diffusivity converges to a $r^{4/3}$ scaling law for $r$ at the low end of the inertial range ($r\sim 5\times10^{-2}L$). Because $\tau_r$ is greater 
for larger $r,$ one expects slower convergence at the upper end. The inset of Fig.~\ref{diffufigure} shows the diffusivity as a function of time for 6 different $r$-values in the range $[3,6.3]\times10^{-2}L$. At late times, $K_L$ increases very slowly and in the inertial range appears to approach at long times a Richardson diffusivity 
$K_L(r,\infty)=k_0 \varepsilon^{1/3}r^{4/3}$ with $k_0=1.47,$ comparable to Kraichnan's closure prediction $k_0=2.00$ \cite{Kraichnan66}. 

\begin{figure}[!t] 
\begin{center}  
\includegraphics[width=\linewidth]{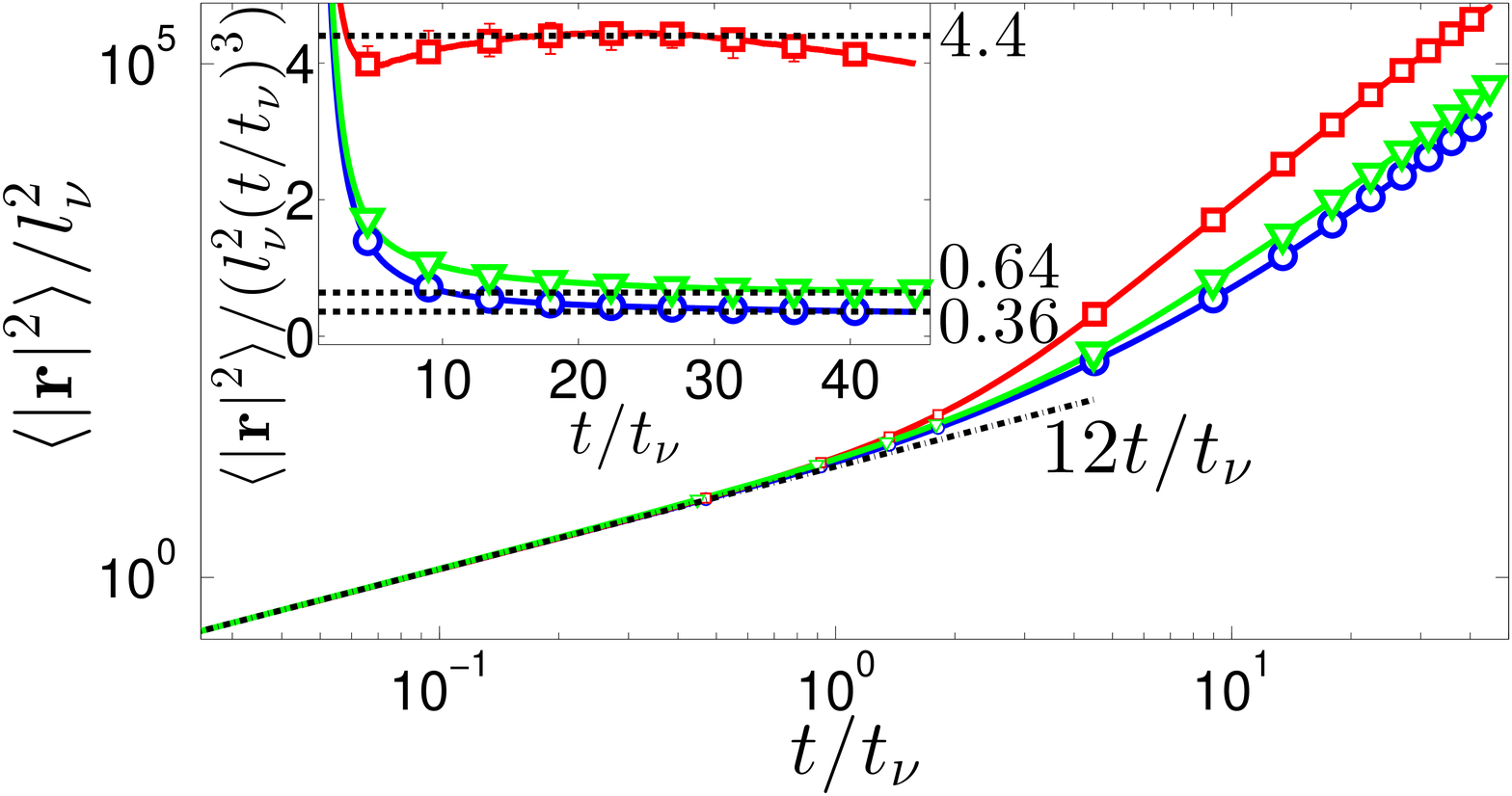}
\caption{Particle dispersion $\langle|\br(t)|^2\rangle$ for the diffusive model  with $K_{ij}^{{\,\!}_{{\,\!}^{{KL}}}}(\br,t)$ ($\textcolor{red}{\square}$), $K_{ij}^{{\,\!}_{{\,\!}^{{KL}}}*}(\br,t)$ ($\textcolor{blue}{\circ}$) 
and the DNS results \cite{Eyink11}($\textcolor{green}{\triangledown}$). The straight dash-dotted line is the fit to the diffusive regime. Inset: curves compensated by $t^3$ (viscous units).} 
\label{evol} 
\end{center} 
\end{figure}

The diffusion model with $K_L(r,t)$ in Fig.~\ref{diffufigure} predicts results for pair dispersion $\langle r^2(t)\rangle$ and PDF $P(r,t)$ 
which may be compared with results from direct numerical simulation (DNS) for the same turbulence database \cite{Eyink11}. To solve the 
diffusion equation we employ a standard Monte Carlo method \cite{ThomsonDevenish05}, using $N=10^5$ samples. 
We first consider pairs separated by various distances $r_0$ at initial time $t_0=0.$ Fig.~\ref{bachelor} for the dispersion 
$\langle |\br(t)-\br_0|^2\rangle$ exhibits a clear Batchelor ballistic regime at  
times $t\ll\tau_{r_0}$. The inset shows convergence toward a $t^3$ regime for times close to $T_L,$ with a Richardson constant $g\simeq 4$. The 
best $t^3$ range occurs for $r_0=4\ell_\nu$ (cf. \cite{Bitaneetal12b}) but there is considerable scatter in the values of $g$ for different $r_0$. 
It was found in \cite{Eyink11} that the Richardson $t^3$-law is more well-defined
for stochastic Lagrangian trajectories solving $d\bx=\bu(\bx,t)dt+\sqrt{2\nu}d\bW(t)$ with an added white-noise, all started at the same initial point. 
For such stochastic trajectories our diffusion model must be  modified (to leading order) by adding $2\nu$ to the diagonal elements of $K_{ij}(\br,t).$ 
Fig.~\ref{evol} plots Monte Carlo results for the dispersion $\langle r^2(t)\rangle$ in this modified diffusion model with $r_0=0,$ together with DNS results of 
\cite{Eyink11}.  The early-time $\sim 12\nu t$-law is reproduced very well by the diffusion model, followed by a reasonable $t^3$ range.
However, the $t^3$ power-law starts too soon and the Richardson constant is $g\simeq4.4\pm 0.2$ (see inset), much larger than the value $g\simeq 0.64$
from DNS \cite{Eyink11}. It is well-known that the KL formula when evaluated by closures \cite{Kraichnan66,Lundgren81} leads to a
value of $g$ which is an order of magnitude too large \cite{OttMann00}. Our results show that this defect is intrinsic to the KL
theory, even when their diffusivity formula (\ref{diffu}) is evaluated by Navier-Stokes solutions, not by uncontrolled closures. 

\begin{figure}[!t] 
\begin{center}  
\includegraphics[width=\linewidth ]{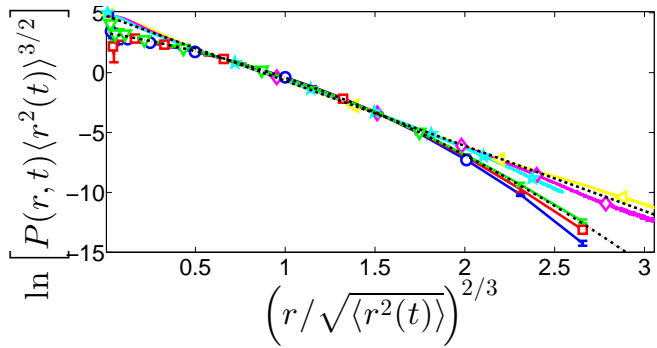}
\caption{The pair separation PDF for our KL diffusion model (DM) and DNS \cite{Eyink11} for 3 different times (with viscous units) in the $t^3$ regime $t=22.37$  (DM: $\textcolor{green}{\triangledown}$, DNS: $\textcolor{yellow}{\triangleleft}$), $t=33.57$ (DM: $\textcolor{red}{\square}$, DNS: $\textcolor{magenta}{\diamond}$)) and $t=44.79$ (DM: $\textcolor{blue}{\circ}$, DNS: $\textcolor{cyan}{\star}$). Infinite Reynolds self-similar 
PDF's are shown for Richardson (straight dashed line) and for KL theory (dot-dashed curve).} 
\label{proba} 
\end{center} 
\end{figure}

To understand why, consider the two main assumptions which led to (\ref{diffu}). The mean-field approximation has no obvious systematic 
effect on the rate of dispersion, but the short-memory approximation must increase the diffusivity. Note indeed that the ratio $P(r,s)/P(r,t)$ 
for $s<t$ in the effective diffusivity $K_{ij}^*(r,t)$ is $<1$ at the peak of $r^4P(r,t)$ where most of the contribution to $\langle r^2(t)\rangle$ arises, 
under the normalization $\int r^2P(r,t)\,dr=1.$ This can be checked for the DNS results of \cite{Eyink11} and it is a simple calculus exercise 
to prove for Richardson's self-similar PDF $P(r,t)=(B/\langle r^2(t)\rangle^{3/2})\exp\left[-A(r/\langle r^2(t)\rangle^{1/2})^{2/3}\right],$ which 
agrees well with the DNS. Thus, setting the ratio $=1$ increases the diffusivity near the peak. To check whether this effect 
can account quantitatively for the excess diffusivity in KL theory, we reintroduce the ratio of PDF's into the KL formula,
using $P$-values from \cite{Eyink11}:
\begin{eqnarray}
K_{ij}^{{\,\!}_{{\,\!}^{{KL}}}*}(\br,t)&=&\int^0_{-t} ds\,\,\langle \delta u_i(\br,0)\delta u_j(s|\br,0)\rangle\frac{P(\br,s)}{P(\br,0)}. \,\,\,\, \label{diffustar}
\end{eqnarray} 
Monte Carlo results with $N=10^5$ for this modified KL diffusivity are also plotted in Fig.~\ref{evol}, showing a $t^3$ regime with a reduced Richardson 
constant $g^*\simeq 0.36\pm0.02$. We conclude that the overestimated dispersion in the KL theory is mainly due to the neglect of memory effects.

Although the short-memory approximation introduces some quantitative errors, it and the other approximations we have made are expected
to be qualitatively correct in the limited range of dispersions $r\simeq \langle r^2(t)\rangle^{1/2}.$ To test this, Fig.~\ref{proba} plots $P(r,t)$ with 
similarity scaling at three different times in the $t^3$ regime for the diffusive model with $K_{ij}^{{\,\!}_{{\,\!}^{{KL}}}}(\br,t)$ (using $N=10^6$), the 
DNS results \cite{Eyink11}, Richardson's self-similar PDF, and the self-similar PDF of KL theory \cite{Kraichnan66,OttMann00}. All the results 
(different models and different times) collapse well in the range $[0.5,1.6]$ of the similarity variable $\rho=(r/\langle r^2(t)\rangle^{1/2})^{2/3},$ 
as expected. It is also true that the DNS results agree better with Richardson's solution over a longer range, while our KL diffusion model results agree best 
with the infinite Reynolds-number KL similarity  solution. Nothing should be concluded about differences at $\rho>3.0,\, 2.1,\, 1.6$ for the three times,
resp., since these lie outside the inertial range.  However, differences inside those ranges must be due to the additional approximation (\ref{SKL}) in KL theory. 
We expect that a diffusion model based instead on (\ref{Sdef3}) should yield a more accurate result.  

The approach developed in this Letter, combining exact relations, physically motivated approximations 
and numerical evaluation, can be exploited also in frameworks that extend Richardson's. An old idea 
\cite{Obukhov59,Lin60a} is to consider the joint transition probability $P(t)=P(\br,\bv,t|\br_0,\bv_0,t_0)$
of both the relative position $\br$ and the relative velocity $\bv$ of two Lagrangian particles, an 
approach which has recently received renewed attention \cite{Bitaneetal12a,Bitaneetal12b}. In this case also 
it is possible to derive exact evolution equations, which can be simplified by rational approximations 
to obtain a simplified equation 
$ \partial_t P +\bv\bdot\grad_r P= \partial_{v^i}\partial_{v^j}\left[ Q_{ij}(t,t_0)P(t)\right] $
with
 \be Q_{ij}(t,s)=\int_{t_0}^t ds\,\, \langle \delta a_i(t|\br,s)\delta a_j(\br,s)|\br,\bv,t;\br_0,\bv_0,t_0\rangle,
 \lb{Qdef} \ee
where $\ba(\bx,t)=(\partial_t+\bu(\bx,t)\bdot\grad)\bu(\bx,t)$ is the Eulerian acceleration field 
and $\ba(t|\bx,s)$ is the corresponding Lagrangian field. In deriving these results, 
a short-memory approximation has been made analogous to that in (\ref{Preleq2}). As long
understood \cite{Obukhov59,Lin60a}, this approximation is justified for a much greater range 
of positions (and velocities) than in (\ref{Preleq2}) because the acceleration field is temporally correlated 
on the scale of the Kolmogorov viscous time $\tau_\nu=(\nu/\varepsilon)^{1/2}$. For times $t-t_0\lesssim\tau_\nu,$
(\ref{Qdef}) implies a velocity ballistic range in which $\langle |\bv(t)-\bv_0|^2\rangle\propto t^2.$
At longer times, it can be expected that there is no dependence of the diffusivity $Q$ upon $\br_0,\bv_0,t_0,$
but numerical evidence suggests that acceleration increments have strong statistical dependence 
upon the instantaneous values $\br,\bv$ of relative positions and velocities \cite{Bitaneetal12a}. This 
implies that the mean-field approximation is more limited in this setting. The formula (\ref{Qdef}) provides a 
systematic framework within which to explore these dependences and to exploit numerical simulation 
data to develop a well-founded model. We are currently pursuing such investigations.

\begin{acknowledgments}
We thank D. J. Thomson for useful discussions. This work was partially supported by NSF grant 
CDI-II: CMMI 0941530 at Johns Hopkins University.
\end{acknowledgments}



\bibliography{DiffuseApprox.bib}

\end{document}